\documentclass{ws-ijmpa}

\begin{document}

\markboth{O.N.~Hartmann}
{Studying Strange Meson Production with FOPI.}

\catchline{}{}{}{}{}
%
%%%%%%%%%%%%%%%%%%%%%%%%%%%%%%%%%%%%%%%%%%%%%%%%%%%%%%%%%%%%%%%%%%%%

\title{Studying Strange Meson Production with FOPI.}

\author{OLAF N.~HARTMANN}

\address{Excellence Cluster Universe, Technische Universit\"{a}t M\"{u}nchen,\\ Boltzmannstr. 2, D-85748 Garching, Germany, and \\
Stefan-Meyer-Institut f\"ur Subatomare Physik, Boltzmanngasse 3, A-1090 Wien, Austria
}

%first\_author@domain\_name}

%\author{SECOND AUTHOR}

%\address{Group, Laboratory, Address\\
%City, State ZIP/Zone, Country\\
%second\_author@domain\_name}

\maketitle

\begin{history}
\received{August 8, 2008}
\revised{\today}
\end{history}

\begin{abstract}

The production of mesons containing strangeness is studied
using elementary and heavy-ion probes with the FOPI detector at GSI-SIS. 
The observed inclusive cross section of neutral kaons points to an
in-medium modification in the production. The momentum dependent ratio of
cross sections of light and heavy target nuclei turns out to be a sensitive
observable for the in-medium potential strength in transport codes. The
experiments will continue studying the production of charged kaons with 
elementary probes including the search for deeply bound antikaon nuclear
clusters in proton proton collisions.

\end{abstract}

\ccode{PACS numbers: 25.80.Hp, 21.65.Jk}

\section*{Introduction.}

\begin{figure}[ht]
\begin{tabular}{ll}
\begin{minipage}{.65\textwidth}
\centerline{\psfig{file=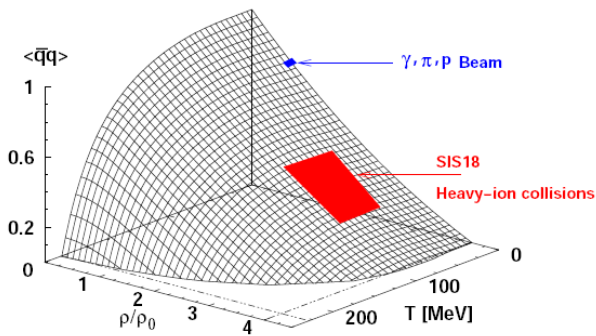,width=\textwidth}}
\vspace*{8pt}
\end{minipage} 

&

\begin{minipage}{.3\textwidth}

\caption{The expectation value of the light quark condensate as a function of temperature and density [1]. The reachable areas in heavy-ion reactions and with elementary probes are indicated. \label{f1}}

\end{minipage}

\end{tabular}

\end{figure}

The present GSI accelerator facility allows for beam energies in the case of heavy ions of up to 2 $A$GeV (for $\frac{A}{q}=2$) and for protons up to 4.7 GeV. Both heavy ion and proton beams can be used to produce a secondary $\pi^+/
\pi^-$ beam which is available with beam momenta from 0.6 to 2.8 GeV/c. In this energy regime the different probes are used to study nuclear matter at high temperatures/high density as well as normal nuclear matter. Fig. \ref{f1} shows the behaviour of the so-called quark condensate $\left< q\overline{q}\right>$ as a function of temperature and density scaled by normal nuclear matter density ($\rho=\rho_0$). The condensate can be read as a measure for the restoration of chiral symmetry; in case of a vanishing condensate chiral symmetry should be restored, while it is broken in the ``real world''. Also a partial restoration of chiral symmetry should have effects to the properties of particles embedded in the appropriate environment. The areas indicated in Fig. \ref{f1} should illustrate the range of $\left< q\overline{q}\right>$ for heavy ion reactions and elementary probes such as proton and pion beams. A sizeable reduction of $\left< q\overline{q}\right>$ is visable. Hence, there might be non-trivial in-medium effects like mass shift, resonance broadening, change in production cross sections and/or the formation of bound states.

The search for experimental evidences for those in-medium effects on hadron properties is part of the FOPI program. Mesons containing strange quarks (charged and neutral kaons, $\phi(1020)$) are of particular interest since at the very moment of the reaction at least one s$\overline{\mbox{s}}$ pair has to be produced. In particular the K$^-$, which contains one $s$ quark, interacts heavily with the surrounding medium (see e.g. [\refcite{kaons1,kaons2}]) and its attractive potential may lead to the formation of bound states.

\section*{The Detector Setup.}

\begin{figure}[ht]
\begin{tabular}{ll}
\begin{minipage}{.5\textwidth}
\centerline{\psfig{file=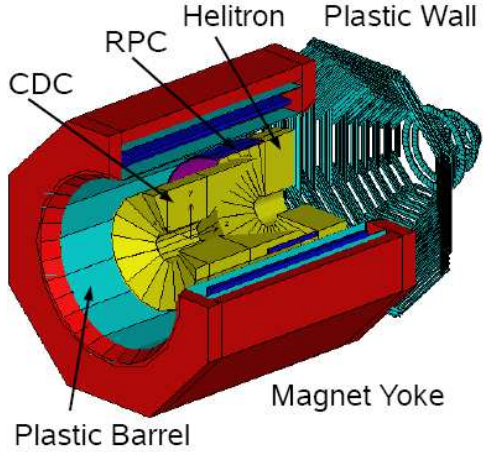,width=\textwidth}}
\vspace*{4pt}
\end{minipage}
&
\begin{minipage}{.45\textwidth}
\vfill
\caption{The FOPI detector setup. The beam is coming from the left lower
corner. The target is surrounded by the Central Drift Chamber (CDC), a plastic barrel for backward angles and a barrel of Resistive Plate Chambers (RPC) for more forward angles. Further forward angles are covered by another drift chamber, the Helitron. The setup is sitting in a solenoidal magnetic field. The Plastic Wall covers the forward hemisphere.
\label{f2}}
\end{minipage}
\end{tabular}
\end{figure}

The FOPI detector (the acronym comes from ``four pi'' in allusion to the solid angle coverage) sits at an external beam line of the SIS accelerator and is a fixed target experiment. Fig. \ref{f2} shows a drawing of the setup.
The target is surrounded by a cylindrical drift chamber (CDC), further out time-of-flight detectors -- plastic scintillators for backward angles, resistive plate counters for more forward angles -- are located as well as an additional radial drift chamber (Helitron) for forward going particles. The whole setup sits in the magnetic field of a superconducting solenoid which provides 0.6 T magnetic field. The forward hemisphere is covered by a plastic wall (time-of-flight).

The RPC barrel [\refcite{as}] has been installed and commisioned recently. With its execellent time-of-flight resolution of less than 100 ps for the full system charged kaons can be identified up to momenta of $\sim$ 1 GeV/c.

\section*{Pion Induced K$^{0}_{\mbox{\tiny S}}$ Production.}

As mentioned in the introduction, GSI offers a secondary beam of charged pions. In a first production experiment with a $\pi^-$ beam FOPI has studied the reaction (\ref{eq1}) in medium [\refcite{lotfi1,lotfi2}]. As target nuclei C, Al, Cu, Sn and Pb where used. 

\begin{equation}
\label{eq1}
1.15 \mbox{ GeV/c } \pi^- + \mbox{p} \rightarrow \mbox{K}^0 + \Lambda
\end{equation}

\begin{figure}[ht]
\begin{tabular}{ll}
\begin{minipage}{.55\textwidth}
\centerline{\psfig{file=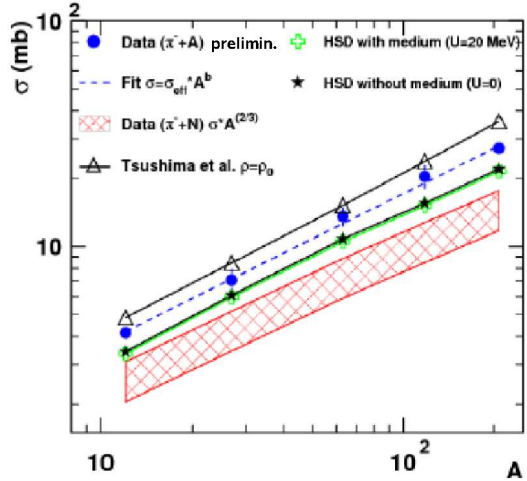,width=\textwidth}}
\vspace*{4pt}
\end{minipage}
&
\begin{minipage}{.4\textwidth}
\caption{The inclusive K$^0$ cross section as a function of the target nucleus $A$. The (blue) full circles are the FOPI data points. The dashed line is the fit (see eq. (\ref{eq2})). The hatched area denotes the expectation scaled from the elementary cross section. The open triangles are the predictions from Tsushima et al., and stars and crosses are predictions from the HSD transport code.\label{kocrosss}}
\end{minipage}
\end{tabular}
\end{figure}

In Ref. \refcite{tsushi} the cross section for reaction (\ref{eq1}) is predicted to be lowered by approximately a factor four when going from $\rho=0$ to $\rho=\rho_0$.
The final state particles of reaction (\ref{eq1}) can be reconstructed in FOPI by calculating the invariant mass of the charged decay particles $\Lambda\rightarrow\mbox{p}+\pi^-$, K$^0_{\mbox{\tiny S}}\rightarrow\pi^+\pi^-$. The background channels, involving $\Sigma$ hyperons and charge exchange reactions, cannot be separated from the final state of reaction (\ref{eq1}). Hence, the inclusive cross section of K$^0$ production is determined, as a function of the target nucleus' mass $A$. The result is shown in the blue full circles in Fig. \ref{kocrosss}.

The data are fitted by a power law using an effective cross section $\sigma_{\mbox{\tiny eff}}$:

\begin{equation}
\label{eq2}
\sigma = \sigma_{\mbox{\tiny eff}}\cdot A^b; \sigma_{\mbox{\tiny eff}}=(0.87\pm0.13)\mbox{mb}, b=0.67\pm0.03
\end{equation}

As written in (\ref{eq2}) the slope parameter is consistent with $\frac{2}{3}$ which indicates that the incoming $\pi^-$ are absorbed at the nucleus' surface. This finding agrees with the mean free path of a $\pi^-$ of $\sim$1 Gev/c in nuclear matter which is about 1 fm.\\
The hatched area in Fig. \ref{kocrosss} is the expected cross section as scaled from the elementary reaction by the found $A$ dependence. The data points ly significantly above this expectation, which points to an in-medium modification of the cross section when going to nuclear targets.\\
The predicted cross sections from Ref. \refcite{tsushi} are represented by the black open triangles in Fig. \ref{kocrosss}. The points overestimate the data, which suggests that the reached density in the pion induced reaction is smaller than $\rho_0$. This is again in agreement with the finding of a surface absorption of the incoming $\pi^-$.

\begin{figure}[ht]
\begin{tabular}{ll}
\begin{minipage}{.6\textwidth}
\centerline{\psfig{file=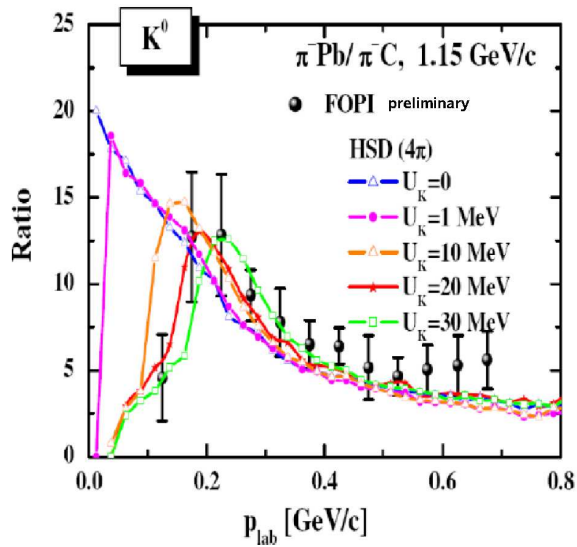,width=\textwidth}}
\vspace*{4pt}
\end{minipage}
&
\begin{minipage}{.35\textwidth}
\caption{The ratio of cross sections from Pb/C as a function of the kaon momentum. The lines represent HSD transport code calculations with different depths of the in-medium potential. \label{ratio}}
\end{minipage}
\end{tabular}
\end{figure}

Finally, the green open crosses and stars show the calculated cross section from the transport model HSD (hadron string dynamics, see Ref. \refcite{hsd}) with and without including a K$^0$ in medium potential, respectively. Both sets of points ly nearly on top of each other, thus the inclusive cross section is not a sensitive observable in that respect. A more promising observable is found in the ratio of cross sections between the heavistest (Pb) and lightest (C) target as a function of the K$^0$ momentum. The data points are shown in Fig. \ref{ratio} in comparison to the predictions from the HSD transport code applying different depths of the in-medium potential. As one can see from the figure, only kaons with low momenta show a sensitivity to the potential depth, hence those who spend more time in the nucleus. The comparison between data and model predictions suggests a potential depth of 20-30 MeV. This finding agrees well with the dataset from the ANKE collaboration (see e.g. Ref. \refcite{rudy}) on K$^+$ production. The cross section ratio shows a similar shape, and the extracted potential depth from the comparison to the CBUU tansport code yields 20 MeV, too.\\
Furthermore, the K$^+$ and K$^-$ in-medium potential could be extracted from the K$^-$/K$^+$ ratio in heavy ion-collisions as a function of the kinetic energy in the center of mass [\refcite{krysiek}]; also here, a K$^+$ in-medium potential of 30 MeV has been extracted which is in good agreement with the latter two.

\section*{Pion Induced K$^+$K$^-$ Production.}
As a continuation of the experiment described in the previous section, FOPI has an accepted experiment proposal to study the elementary and in-medium properties of the reaction (\ref{reac2}).

\begin{equation}
1.7 \mbox{ GeV/c } \pi^- + \mbox{p} \left( \rightarrow \phi + \mbox{n} \right) \rightarrow \mbox{K}^+\mbox{K}^- + \mbox{n}
\label{reac2}
\end{equation}

The resonant (i.e. via producing the $\phi(1020)$ meson) and the non-resonant production of K$^+$K$^-$ pairs with a $\pi^-$ beam on liquid hydrogen, carbon and lead will be studied. The change in $\phi$ production will be measured in form of the transparency ratio $T_Z$ (\ref{eq4}).

\begin{equation}
T_Z = \frac{\sigma_{\pi^- A\rightarrow\phi X}}{Z^{\alpha}\sigma_{\pi^- p\rightarrow\phi X}}
\label{eq4}
\end{equation}

$T_Z$ represents the ratio of the measured cross section on a nucleus $A$ to the cross section of the elementary process scaled by the number of protons in the target nucleus and a factor which accounts for surface effects like already seen in (\ref{eq2}). The analysis of this quantity will be done in analogy to Ref. \refcite{momue}, at the end the $\phi$N cross section can be determined. The photoproduction experiment described in Ref. \refcite{leps} reported already a significant enhancement of this cross section in medium as expected from its vacuum value.\\
At the same time, the co-production of K$^+$K$^-$ can be investigated by taking those pairs whose invariant mass does not coincide with the $\phi$ resonance. 

\begin{figure}[ht]
\begin{tabular}{ll}
\begin{minipage}{.65\textwidth}
\centerline{\psfig{file=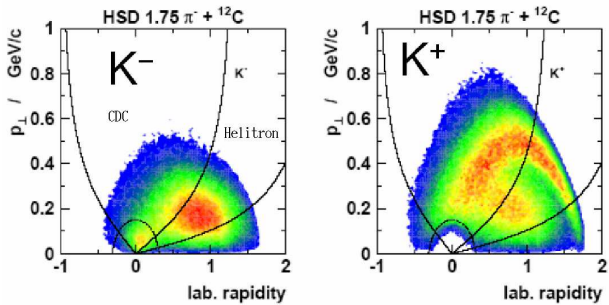,width=\textwidth}}
\vspace*{4pt}
\end{minipage}
&
\begin{minipage}{.3\textwidth}
\caption{The phase space distribution of co-produced K$^-$ (left) and K$^+$ right calculated with the HSD transport code. The lines indicated the FOPI detector acceptance. \label{hsd_kpm}}
\end{minipage}
\end{tabular}
\end{figure}

Fig. \ref{hsd_kpm} shows a calculation done with the HSD transport code for the reaction 1.75 GeV/c $\pi^-$ + $^{12}$C. The phase space distributions in transverse momentum versus rapidity (lab frame) for K$^-$ and K$^+$ are plotted. The solid lines correspond to constant polar angles (lab frame) and show the acceptance of the drift chambers of FOPI (i.e. the region of phase space where particle identification is possible), the dashed line denotes the particles which probably have too low energy to reach the drift chambers. As one can see, the phase space is nicely covered by th FOPI acceptance.
One can guess from the two distributions in Fig. \ref{hsd_kpm} that the spectral shape comparison of K$^+$ to K$^-$ provides information on the in-medium potential, especially interesting for the K$^-$.

\section*{Search for Deeply Bound Antikaonic Nuclear Clusters.} 

The search for the possible existence of deeply bound clusters of antikaons with nucleons receives particular attention. In 2002, the authors of Ref. \refcite{ay1} predicted the existence of antikaon nuclear bound states with remarkable large binding energies and relatively small widths. The lighest of these clusters, formed by one K$^-$ and two protons (here denoted as [K$^-$pp]), is expected to have a mass of 2223 MeV/c$^2$ (i.e. below the sheer sum of the three hadrons), a binding energy of 48 MeV and a width of 61 MeV. The same authors also propose an enhanced probability to form this state in proton proton collisions [\refcite{ay2}].

\begin{equation}
\mbox{p} + \mbox{p} \rightarrow \mbox{K}^+  + \Lambda^{*} + \mbox{p}, Q \approx 1.6 \mbox{ GeV/c}
\label{eq5}
\end{equation}

Reaction (\ref{eq5}) is suggested to yield [K$^-$pp] via the $\Lambda^{*}$p doorway state. The maximum cross section for this process is predicted to be reached around 3 GeV proton energy and to be in the order of a few microbarns (compared to 44 mb total cross section in p+p). An efficient trigger to enrich experimentally the interesting events is being built and tested at the moment.

\begin{figure}[ht]
\begin{tabular}{ll}
\begin{minipage}{.5\textwidth}
\centerline{\psfig{file=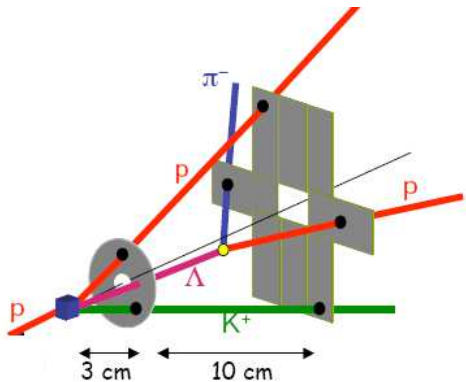,width=\textwidth}}
\vspace*{4pt}
\end{minipage}
&
\begin{minipage}{.45\textwidth}
\caption{Schematical drawing of the $\Lambda$ trigger made by two layers of silicon strip detectors. The trigger decision is based on the charged particle multiplicity. In the drawn case, a $\Lambda$ hyperon decays in between the two detector layers, and the second layers sees a multiplicity increased by two.
\label{silvioschem}}
\end{minipage}
\end{tabular}
\end{figure}

The decay of [K$^-$pp] gives a hyperon and a proton in the final state. The $\Lambda$ hyperon decays with a branching ratio of 64\% into p + $\pi^-$ (\ref{eq6}).

\begin{equation}
[\mbox{K}^-\mbox{pp}]\rightarrow\Lambda + \mbox{p}\rightarrow \mbox{p} + \pi^- + \mbox{p}
\label{eq6}
\end{equation}

Analysing the invariant masses of $\pi^-$ + p and $\Lambda$ + p from (\ref{eq6}) together with the K$^+$ missing mass from (\ref{eq5}), a [K$^-$pp] cluster like predicted in Ref. \refcite{ay1} could be identified.\\
As mentioned above, the FOPI apparatus will be supplemented by an additional $\Lambda$ trigger which is schematically shown in Fig. \ref{silvioschem}. The working principle of the $\Lambda$ trigger is based on the on-line comparison of charged particle multiplicities. Therefore, two layers of silicon strip detectors are mounted perpendicular to the beam axis at a distance of 3 cm and 13 cm from the target, respectively. If a $\Lambda$ hyperon, e.g. from reaction (\ref{eq6}), decays in between the two layers and its charged decay products are detected in the second layer, the multiplicity jump of +2 can be used for a trigger decision.\\
The appropriate detectors, an annular one, single-sided, divided in 32 sectors, and a patchwork of eight rectangular double-sided (6 cm$\cdot$4 cm, with 16 long strips on one side and 60 short strips on the other side) silicon strip detectors of 1 mm thickness are assembled and equipped with frontend and readout electronics. Special modules are used to provide a fast multiplicity signal.\\[1ex]
A setup of two silicon strip detectors (square area) has been tested in-beam. The trigger concept showed to work with a very good trigger purity [\refcite{martin}].\\
The final experiment, scheduled for the year 2009, will use the FOPI detector together with the $\Lambda$ trigger (like shown in Fig. \ref{silvioschem}). The target will be a cell filled with liquid hydrogen. In addition, a new start detector capable to deliver a fast timing signal for the minimal ionizing proton beam particles as well as a beam profile monitor and a veto detector to ensure that the target is hit in its center are constructed. In september 2008 a final in-beam test of all components will take place.

\begin{figure}[ht]
\centerline{\psfig{file=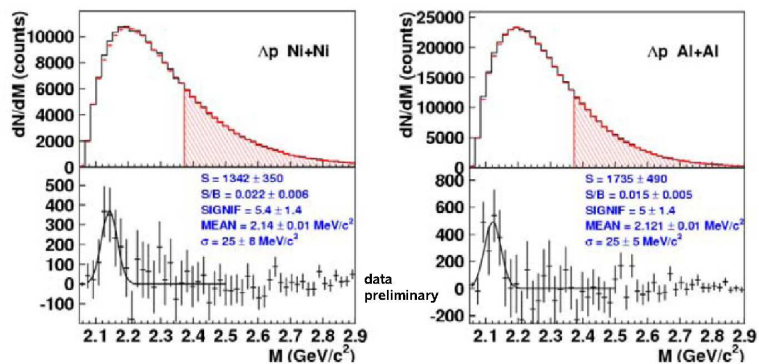,width=.8\textwidth}}
\vspace*{4pt}
\caption{Invariant mass spectra from $\Lambda$p in heavy-ion reactions (left: Ni+Ni, right: Al+Al). The top panel shows the spectra together with the mixed event background (normalized in the hatched area). The bottom panel shows the background subtracted signal.\label{hic}}
\end{figure}

\section*{Outlook}
Besides the two experiments with $\pi^-$ and proton beams discussed in the previous sections, FOPI pursues an experimental program with heavy-ion collisions. \\
Correlations of $\Lambda$p pairs, like from reaction (\ref{eq6}), are investigated in heavy ion reactions, too. Fig. \ref{hic} shows the  $\Lambda$p invariant mass distributions of the two systems Ni+Ni and Al+Al. An excess at a mass of $\sim$2.14 GeV/c$^2$ is observed with an statistical signifcance of $\sim$5. The interpretation of this observation is, however, not straight forward, as it might come from a bound state, final state interaction or be a partial invariant mass of a heavier state like $^4_\Lambda$He. The search for strange clusters is continued (correlations of $\Lambda$ with protons, deuterons and tritons) as well as the search for other multibaryonic states (H$_1^+$).\\
Furthermore, the production of $\Lambda$, K$^0$ and $\phi$ mesons as well as resonances like $\Sigma^{*}$(1385) and K$^*$(892) is studied (see e.g. [\refcite{xavier}]).\\

\section*{Acknowledgments}
The collaboration acknowledges the support by the German Bundesministerium f\"ur Bildung und Forschung, the Helmholtz-Gemeinschaft Deutscher Forschungszentren and the DFG-Excellence Cluster Universe-TU M\"{u}nchen.

%\begin{thebibliography}{000} %for 3 digits

\end{document}